\documentclass[aps,prl,twocolumn,groupedaddress,preprintnumbers,showpacs,amsmath,amssymb]{revtex4}

\usepackage{bm}

\begin{document}

\preprint{Submitted to Phys. Rev. Lett.}

\title{Linear Stability and Subcritical Turbulence in Rotating
Shear Flows}


\author{Pierre-Yves Longaretti}
\email{pyl@obs.ujf-grenoble.fr}
\homepage{http://www-laog.obs.ujf-grenoble.fr/~pyl/}
\affiliation{Laboratoire d'Astrophysique de Grenoble, BP 53X,
F-38041, Grenoble, France}


\date{03/20/03}

\begin{abstract}

The relation between rotating plane Couette and Taylor-Couette
flows is clarified. Experimental data are used to quantify the
behavior of the minimum Reynolds number for subcritical turbulence
as a function of rotation and curvature. This last dependence is
understood through a phenomenological analysis, which also allows
us to relate the subcritical turbulent transport efficiency to the
transition Reynolds number. This implies that the Coriolis force
reduces the efficiency of subcritical turbulent transport with
respect to nonrotating flows, and resolves an ongoing controversy.

\end{abstract}



\maketitle

Shear flows constitute one of the prototypical type of
hydrodynamical flows. Furthermore, they are commonly found in
various instances, e.g. geophysical and astrophysical contexts.
This makes the understanding of their properties an important
issue from both a fundamental and a practical point of view. Most
prominently, characterizing turbulence and turbulent transport in
such flows is critically needed, as astrophysical and geophysical
shear flows are usually fully turbulent because of the large
scales involved.

A large body of experimental evidence has been collected on
Taylor-Couette flows in the linearly unstable regime (see, e.g.,
Ref.~\cite{And86} and references therein). Much less is known
about the linearly stable, or subcritical, regime
\cite{Tay36,Wen33,Rich01}. Plane Couette flows (rotating or not)
are more difficult to realize experimentally, and have been less
extensively studied \cite{Till92,TA96,Dauch94}.

On the theoretical side, intense efforts have been devoted to the
understanding of turbulence spectral and statistical properties
(see, e.g., \cite{les} for an introduction to the subject); also,
mostly for practical purposes, rather complex turbulent transport
models have been developed \cite{Spe91}. Unfortunately, in spite
of these remarkable successes, some basic properties such as
magnitude of the critical Reynolds number of fully developed
turbulence, or its dependence on rotation, are not yet understood.
Worse, even the most sophisticated Reynolds stress closure models
fail to account for the existence of subcritical turbulence in the
presence of a stabilizing rotation.

Some of these shortcomings are addressed in the present
investigation, mostly through a phenomenological analysis of the
effects of rotation and curvature in rotating plane Couette flows
and Taylor-Couette flows. This provides us with an understanding
of some of the characteristic features of subcritical shear
turbulence, and, most importantly from a practical point of view,
establishes a relation between the efficiency of subcritical
turbulent transport and the magnitude of the minimum turbulent
Reynolds number. The emphasis on subcritical turbulence follows
for the following reasons. First, an analysis of angular momentum
transport suggests that both linearly stable and unstable fully
turbulent Taylor-Couette flows are controlled by similar nonlinear
physics \cite{Tay36,RZ99}. Secondly, it turns out that subcritical
shear flows are easier to analyze from a phenomenological point of
view, and they are in any case important in themselves.

Let us first reexamine the connection between rotating plane
Couette flows and Taylor-Couette flows. The Navier-Stokes equation
for rotating plane Couette flows reads, in the rotating frame

\begin{equation}\label{NSR}
  \frac{\partial{\bm w}}{\partial t}+{\bm w}.{\bm \nabla}{\bm w}
  =-\frac{{\bm\nabla}{P^*}}{\rho}-2{\bm \Omega}\times{\bm w}+\nu\Delta{\bm w}.
\end{equation}

\noindent In this equation, $P^*$ represents the sum of the
inertial term and of the gas pressure term, as usual when
considering incompressible flows. Such flows are characterized by
two dimensionless numbers, the Reynolds number $Re= \Delta V\Delta
L/\nu\sim |{\bm w}.{\bm\nabla}{\bm w}/\nu\Delta{\bm w}|$, and a
rotation (inverse Rossby) number $R_\Omega=\epsilon 2\Omega\Delta
L/\Delta V\sim |2{\bm \Omega}\times{\bm w}/{\bm w}.{\bm
\nabla}{\bm w}|$, which is a global measure of the rotation
parameter defined by $S\equiv -2\Omega/(dV/dy)$. In these
definitions, the $x$ axis is assumed to lie in the streamwise
direction, the $y$ axis in the shearwise direction, and the $z$
axis in the spanwise direction (rotation axis); $V(y)$ is the mean
flow (along $x$). The rotation number is positive (resp. negative)
($\epsilon=\pm 1$) depending on whether the rotation is cyclonic
(resp. anticyclonic).

Stability limits for inviscid rotating plane Couette flows can be
determined through a displaced particle analysis
\cite{Trit92,Tritdav81}. Indeed, although the total work of the
Coriolis force vanishes, the work of the Coriolis force component
in the streamwise direction during a fluid particle displacement
in the shearwise direction does not. Comparing the resulting
post-displacement streamwise velocity to the equilibrium one
implies that the flow is linearly unstable if $-1<S<0$, and
linearly stable otherwise, as the squared oscillation frequency of
fluid particles is given by $(dV/dy)^2 S(S+1)$.

The Navier-Stokes equation for Taylor-Couette flows is most
meaningfully compared to that of rotating plane Couette flows when
it is dispelled in a frame rotating with some characteristic
rotation velocity of the flow $\Omega_o$ (e.g., the average
angular velocity of the two cylinders), as only differential
rotation plays a role in the generation of turbulence. Designating
by ${\bm w}$ and $\phi$ the velocity and azimuthal coordinate in
the rotating frame, the Navier-Stokes equation for ${\bm w}=(w_r,\
w_\phi,\ w_z)$ becomes

\begin{align}\label{NSCT}
  \frac{\partial{\bm w}}{\partial t}+ {\bm w}.{\bm\nabla'}{\bm w} &
  + 2{\bm \Omega_o}\times{\bm w} - \nonumber\\ & \frac{w_{\phi}^2}{r}{\bm e}_r
  +\frac{2w_{\phi}w_r}{r}{\bm e}_{\phi}=
  -\frac{\bm\nabla P^*}{\rho} +
  \nu\Delta{\bm w},
\end{align}

\noindent where $P^*$ includes again all gradient terms, and

\begin{align}\label{nabla'}
{\bm w}.{\bm\nabla'}{\bm w}\equiv({\bm w}.{\bm\nabla} w_r) {\bm
e}_r+ [r{\bm w}.{\bm\nabla} (w_{\phi}&/r)] {\bm e}_{\phi} + \nonumber\\
& ({\bm w}.{\bm\nabla} w_z) {\bm e}_z.
\end{align}

\noindent With this new definition, the contribution of the
azimuthal velocity derivative to the ``advection" term vanishes
when the fluid is not sheared; as a consequence, the ${\bm
w}.{\bm\nabla'}{\bf w}$ can appropriately be named the
``advection-shear" term. This reexpression of the Navier-Stokes
equation is essential: explicitly isolating shear terms in the
advection term distinguishes the effect of the shear, upon which
the Reynolds number is built, from the others, the flow curvature
and rotation. Indeed, with respect to rotating plane Couette
flows, Eq.~(\ref{NSR}), two new terms are added, $w_{\theta}^2/r$
and $2w_rw_\theta/r$, which represent the effect of the curvature
of the flow due to the cylindrical geometry. As will soon become
apparent, Eq.~(\ref{NSCT}) is extremely helpful to develop a
correct physical analysis of Taylor-Couette flows.

This reexpression of the Navier-Stokes equation shows that the
three dimensionless numbers which best characterize Taylor-Couette
flows are the Reynolds number $Re= R\Delta\Omega\Delta R/\nu\sim
|{\bm w}.{\bm\nabla'}{\bm w}/\nu\Delta{\bm w}|$, the rotation
number $R_\Omega=\epsilon 2\Omega\Delta R/R\Delta\Omega\sim |2{\bm
\Omega_o}\times{\bm w}/{\bm w}.{\bm\nabla'}{\bm w}|$, which is a
global measure of the rotation parameter $S=2\Omega/(rd\Omega/dr)$
\footnote{The change of sign in the definition of $S$ results from
the proper identification of the various axes between rotation
plane Couette flows and Taylor-Couette flows.}, and the curvature
number $R_c=\Delta R/R\sim {\bm w}^2/(r|{\bm w}.{\bm\nabla'}{\bm
w}|$, which is a global measure of the curvature parameter $C=
2w_\theta/(r^2d\Omega/dr)$. As will become apparent shortly, these
numbers are to be preferred over the most widely used inner and
outer Reynolds number, and ratio of cylinder radii $\eta$.

Note also that rotating plane Couette flows can be viewed as a
limiting case of Taylor-Couette flows, when $R_c\rightarrow 0$ at
finite $R_\Omega$. This implies that there must be a connection
between the stability limit of rotating plane Couette flows
recalled above, and the Rayleigh criterion. Indeed, the displaced
particle analysis presented above is easily transposed to
Taylor-Couette flows \cite{long03}. Let us consider a fluid
particle at some arbitrary radius $r_o$, and chose $\Omega_o =
\Omega(r_o)$. With this choice, the curvature terms are second
order with respect to the radial displacement of the particle, and
can be neglected. Therefore, the argument for rotating plane
Couette flows applies in exactly the same way to Taylor-Couette
flows, and implies that they are linearly unstable when $-1 <
S\equiv 2\Omega/(rd\Omega/dr) < 0$, where $S$ is evaluated at
$r=r_o$. Equivalently, the restoring (or destabilizing) frequency
reduces to the well-known epicyclic frequency: $(rd\Omega/dr)^2
S(S+1)=(2\Omega/r)d(r^2\Omega)/dr$. This implies that stability
follows when the Rayleigh criterion is satisfied, as expected, and
shows that this criterion can be reinterpreted in terms of the
action of the Coriolis force in the rotating frame.

A similar relation exists between the two types of flows in what
concerns the minimum Reynolds number for the onset of subcritical
turbulence. However, this quantity for Taylor-Couette flows
depends both on rotation number and curvature number, an important
feature whose meaning and consequences have apparently escaped
attention until now.

Transition to turbulence in subcritical flows occurs abruptly,
without undergoing the series of bifurcations which is
characteristic of linearly unstable flows. Note however that the
minimum turbulent Reynolds number is not defined in a clear-cut
way, as the transition from laminar to turbulent occurs in general
at higher Reynolds number than the transition from turbulent to
laminar, and as there is a range of Reynolds numbers over which
the flow changes from highly intermittent to fully turbulent.
Nevertheless, the data quoted here are more or less directly
comparable, as they are rather characteristic of fully turbulent
flows, except for the data of Ref.~\cite{Rich01}, which are
therefore used in a qualitative rather than quantitative way in
the discussion, when needed.

The available data on cyclonically rotating plane Couette flows
imply that the minimum turbulent Reynolds number increases with
the rotation number as \cite{TA96}

\begin{equation}\label{ReRC}
Re_m\simeq Re_c+ a_c R_\Omega,
\end{equation}

\noindent where $Re_c\simeq 1400$ is the transition Reynolds
number of plane Couette flows \cite{Till92}, and $a_c\simeq
26000$. The data were obtained for low cyclonic ($0.1 \gtrsim
R_\Omega$) rotation.

The experimental results of Wendt \cite{Wen33} and Taylor
\cite{Tay36} for Taylor-Couette flows in cyclonic rotation were
obtained by maintaining the inner cylinder at rest. In this case,
$R_c=R_\Omega=\Delta R/R$, which makes it a priori difficult to
distinguish the effects of rotation from those of curvature.
Nevertheless, for $\Delta R/R \lesssim 1/20$, the data are clearly
consistent with Eq.~(\ref{ReRC}) (in agreement with the existence
of the rotating plane Couette flow limit for Taylor-Couette
flows), while for $\Delta R/R \gtrsim 1/20$, the minimum Reynolds
number varies as $Re_m\simeq Re^*(\Delta R/R)^2$, with $Re^*\simeq
6\times 10^5$ \cite{RZ99}. This last behavior is necessarily due
to the effect of the flow curvature, and not rotation. This
follows first because the predictions of Eq.~(\ref{ReRC}) fall way
below the required minimum turbulent Reynolds number when
extrapolated to larger $\Delta R/R$, and because the experiments
of Wendt for marginally stable anticyclonic flows are globally
consistent with the quadratic dependence of $Re_m$ for large
enough $\Delta R/R$ \cite{long03}; also, the results of
Ref.~\cite{Rich01} are in qualitative agreement with this
conclusion \cite{long03}. All these arguments imply that, for
cyclonically rotating Taylor-Couette flows,

\begin{equation}\label{Remcycl}
  Re_m\simeq Re_c + a_{c} R_\Omega + Re^* R_c^2.
\end{equation}

This relation applies to moderate rotation rates ($Ro \lesssim
1/2$), where the extrapolation of Eq.~(\ref{ReRC}) is still valid
and the turbulence still three-dimensional. The fact that the
rotation and curvature contributions add is implied by the data of
Richard \cite{Rich01} (see Ref. \cite{long03} for details). Notice
that the contributions of either rotation ($a_c R_\Omega$) or
curvature ($R_e^* R_c^2$) become comparable to $Re_c$ when the
rotation or curvature number is $\simeq 1/20$. This remarkable
feature is no coincidence. Indeed, the turbulence regeneration
mechanism identified in nonrotating plane Couette flows
\cite{Ham95,Wal97} has an overall timescale $t_{turb}\simeq 100
(dU/dy)^{-1}\simeq 100 (rd\Omega/dr)^{-1}$. The timescale
associated with the Coriolis term is $t_\Omega\simeq\Omega$, while
the timescale associated with the curvature terms is $t_curv\simeq
r/w_\theta\simeq\Delta\Omega^{-1}$. Therefore, these effects are
expected to significantly affect the turbulence regeneration
mechanism when these timescales decrease to become comparable to a
few times $t_{turb}$, or equivalently, when $Ro$ or $R_c$ exceed a
few percents. This physical constraint is what primarily
determines the magnitude of $a_c$ and $Re^*$ (once the form of the
dependence on $R_\Omega$ and $R_c$ is known). Indeed, for $R_c,\
R_\Omega \gtrsim 1/20$, requiring $a_c R_\Omega,\ Re^* R_c^2
\gtrsim Re_c$ implies that $a_c \sim 10^4$ and $Re^* \sim 10^5$.

The explanation of the dependence of $Re_m$ on $R_c$ of Zeldovich
\cite{Zel81} is inconsistent with some of the data \cite{RZ99},
while the explanation of Dubrulle \cite{Bulle93} is incompatible
with the fact that this is an effect of the flow curvature. In
fact, it can be understood by extending the turbulent viscosity
description \cite{long02}, an argument which is clarified and
noticeably improved here. Characterizing turbulent fluctuations
responsible for turbulent transport by their coherence scale $l_M$
and velocity fluctuation $v_M$, a simple analogy with molecular
transport implies that \cite{Pr25} $\langle\delta v_r \delta
v_\phi\rangle \simeq \nu_t r d\langle\Omega\rangle/dr$, with
$\nu_t\simeq l_M v_M$. In a Kolmogorov cascade picture, the rate
of energy dissipation $\epsilon$ can be expressed in terms of the
same quantities as $\epsilon\simeq C^* v_M^3/l_M \simeq (\nu_t/2)
(r d\langle\Omega\rangle/dr)^2$, where $C^*\simeq 0.1$ is a rather
universal constant \cite{Spe91}. In this picture, the turbulence
regeneration mechanism involves scales in the range $\Delta L$ to
$l_M$, which do not dominate the turbulent transport; furthermore,
anisotropy is neglected for the moderate shear and rotation
numbers involved. Finally, a flow submitted to a velocity shear
tries to restore global thermodynamic equilibrium by suppressing
the shear and transporting momentum in the shearwise direction. A
linearly stable system has only two means to achieve this purpose,
laminar or subcritical turbulent transport, and will chose the
most efficient one. Turbulent transport is dominant when
$|\langle\delta v_r\delta v_\phi\rangle| \gtrsim \nu
|rd\Omega/dr|$, so that, at the transition to turbulence, $\nu_t =
C_\nu \nu$ where $C_\nu$ is a constant of order unity. Taken
together, these relations imply that

\begin{equation}\label{vm}
v_M\simeq
\left(\frac{C_\nu}{(2C^*)^{1/2}}\right)^{1/2}\frac{r\Delta\Omega}{Re_m^{1/2}},
\end{equation}

\begin{equation}\label{lm}
l_M\simeq \left(C_\nu(2C^*)^{1/2}\right)^{1/2}\frac{\Delta
r}{Re_m^{1/2}},
\end{equation}

\noindent where $\Delta r$ is the typical scale of the shear away
from the boundaries, and $r\Delta\Omega$ the corresponding shear
amplitude.

Let us focus for the time being on Taylor-Couette systems
dominated by the curvature. By making the outer cylinder radius
arbitrarily large, this last relation implies that, if $Re_m$ were
constant (independent of the relative gap width), the turbulent
scales could become arbitrarily larger than the radius, at any
given radial location in the flow, which makes no sense.
Reversely, this shows that the local radius $r$, at any flow
location, must impose a limiting scale. Indeed, by imposing
$l_M=\gamma r$ (where the proportionality constant $\gamma$ will
soon be determined), Eq.~(\ref{lm}) is equivalent to \footnote{For
the values of $\Delta R/R$ characteristic of the available
experiments, there is little difference between the global mean
radius $R$ and local values $r$.}

\begin{equation}\label{Re*}
Re_m \simeq \frac{C_\nu(2C^*)^{1/2}}{\gamma^2}\left(\frac{\Delta
r}{r}\right)^2\equiv Re^*\left(\frac{\Delta r}{r}\right)^2.
\end{equation}

In the same vein, the resulting turbulent velocity expression
$\nu_t\simeq (C_\nu/Re^*) r^3 |\Delta\Omega/\Delta r|\equiv \beta
r^3 d\Omega/dr$ also implies that in this regime, turbulence is
controlled by local rather than global conditions in the bulk of
the flow. This turbulent viscosity prescription has indeed been
derived from the data of Wendt \cite {RZ99}, which imply that
$\beta\simeq 10^{-5}$, so that $C_\nu\simeq 6$ and $\gamma\simeq
1/500$.

If turbulence in the bulk of the flow is controlled by local
conditions, this implies that the width of the flow plays no role
in the curvature dominated regime (except indirectly through the
way global conditions might control local ones). This surprising
conclusion can be understood in the following way. For narrow gap
widths (still neglecting the effect of rotation), the flow reduces
to a (nonrotating) plane Couette flow, and turbulence occurs as
soon as $R\Delta\Omega \Delta R/\nu$ exceeds $Re_c$. When $\Delta
R \gtrsim R/20$, the mean radius $R$ locally defines radial
``boxes" of size $R/20$ which are turbulent if
$(rd\Omega/dr)(R/20)^2/\nu \gtrsim Re_c$, so that, for the shear
to be large enough on scale $R/20$, the Reynolds number must
increase quadratically with the gap width. Incidentally, this
shows that the turbulence regeneration mechanism in curvature
dominated flows must be closely related to the one that operates
in nonrotating plane Couette flows. Consequently, the mechanism
identified in Ref.~\cite{Ham95,Wal97} is probably more than a
near-wall turbulence regeneration mechanism. Note in this respect
that this mechanism is largely insensitive to the actual nature of
the boundary condition; only a box size is needed.

What has been achieved so far ? The magnitude of $Re_c$ is
explained by the regeneration mechanism just mentioned, while the
magnitudes of $a_c$ and $Re^*$ have been constrained from the
overall time scale of this mechanism; the meaning of the quadratic
dependence on $R_c$ has also been brought to light. Only the
linear dependence of $Re_m$ on $R_\Omega$ still needs to be
explained. As the Coriolis force term does not pinpoint a
particular length scale (only a time scale is involved), in
contrast to the curvature term, the flow rotation is expected to
affect the regeneration mechanism in an important way, so that
this behavior is beyond the scope of the phenomenological analysis
developed here. It is essential to note that the turbulent
transport $\nu_t\simeq l_M v_M$ based on Eqs.~(\ref{vm}) and
(\ref{lm}) is valid for all $Re \geqslant Re_m$, as long as the
turbulence regeneration mechanism does not change with increasing
Reynolds number; this follows because increasing the Reynolds
number only increases the range of the inertial spectrum, while
the regeneration mechanism necessarily acts on larger scales
\cite{long02,long03}. This feature always allows us to relate the
minimum Reynolds number to the transport efficiency through
Eqs.~(\ref{vm}) and (\ref{lm}) and the turbulent viscosity
prescription, an essential feature from a practical point of view.

Finally, the dependence of $Re_m$ on $R_\Omega$ allows us to
resolve an important controversy on the relevance of subcritical
shear turbulence to accretion disk transport. Most astrophysicists
now believe that the Coriolis force relaminarizes shear flows in
the anticyclonic regime (in particular in the ``keplerian",
$R_\Omega=-4/3$, regime), on the basis of two sets of simulations
\cite{BHS96,HBW99} performed for rotating plane Couette flows with
pseudo-periodic boundary conditions and with a tidal term included
in the generalized pressure $P^*$. On the other hand, a minority
insists that this point of view is inconsistent with the
experimental evidence \cite{RZ99}. This last claim is indeed
correct \cite{Rich01,long03}; nevertheless, this can be reconciled
with the simulation results as is now argued with a much more
focused argument than the one given in Ref.\ \cite{long02}. First,
although no usable data are available in the relevant rotation
dominated anticyclonic regime, the approximate ``Richardson
similarity" of rotating flows \cite{Spe89} (valid for the low
enough rotation numbers of interest here) implies that the minimum
Reynolds number of anticyclonic flows ($R_\Omega \leqslant -1$)
behaves as $Re_m\simeq Re_c + a_{ac}|1+R_\Omega|$ with $a_{ac}
\gtrsim a_c$ (see Ref.~\cite{long03} for details). This implies
that the turbulent amplitude decreases steeply with the rotation
number, as implied by the simulations performed close to marginal
stability \cite{BHS96}. Furthermore, for $R_\Omega= -4/3$, $\Delta
R/l_M\sim (\Delta V)/v_M\gtrsim$ a few $100$, and $\langle \delta
v_r \delta v_\phi\rangle \sim 10^{-3}(\Omega\Delta R)^2$. These
figures imply that the largest simulations of keplerian flows
performed to date \cite{HBW99} miss the mark by lack of
resolution, but not by much.

The numerical results were most probably not recognized as
characteristic of turbulence, because it is widely believed that
turbulent scales and velocity fluctuations are comparable to the
scale of the flow, and mean velocity difference on this scale. On
the contrary, the analysis presented here implies that rotation
reduces the transport generating velocity fluctuations with
respect to nonrotating flows. A fuller account of the relevance of
these features to accretion disk turbulent transport will be given
elsewhere \cite{long03}.






\begin{thebibliography}{}

\bibitem{And86} C. D. Andereck, S. S. Liu, and H. L.
Swinney, J. Fluid Mech., 164, 155 (1986).

\bibitem{Tay36} G. I. Taylor, Proc. Roy. Soc.
London, A 223, 289 (1936).

\bibitem{Wen33} F. Wendt, Ing. Arch., 4, 577 (1933).

\bibitem{Rich01} D. Richard. {\it
Instabilit\'es hydrodynamiques dans les \'ecoulements en rotation
diff\'erentielle.} PhD thesis. Universit\'e de Paris VII (2001).

\bibitem{Till92} N. Tillmark, and
P. H. Alfredsson, J. Fluid Mech., 235, 89 (1992).

\bibitem{TA96} N. Tillmark, and
P. H. Alfredsson. In {\it Advances in Turbulence VI.} (eds,
Gavrilakis {\it et al.}); p. 391. Kluwer (1996).

\bibitem{Dauch94} O. Dauchot,
and F. Daviaud, Phys. Fluids, 7, 335 (1994).

\bibitem{les} M. Lesieur, Turbulence in Fluids, Kluwer Academic
Press (1987).

\bibitem{Spe91} C. G. Speziale, Ann. Rev. Fluid Mech., 23, 107
(1991).

\bibitem{RZ99} D. Richard, and J.-P. Zahn,
Astron. Astrophys., 347, 734 (1999).

\bibitem{Trit92} D. J. Tritton, J. Fluid
Mech., 241, 503 (1992).

\bibitem{Tritdav81} D. J. Tritton, and P. A. Davies.
In {\it Hydrodynamic Instabilities and the Transition to
Turbulence} (ed. Swinney and Gollub), p. 229. Springer (1981).

\bibitem{long03} P.-Y. Longaretti. {\it In preparation.}

\bibitem{Ham95} J. H. Hamilton, J. Kim, and F. Waleffe, J. Fluid Mech., 287,
317 (1995).

\bibitem{Wal97} F. Waleffe, Phys. Fluids, 9,
883 (1997).

\bibitem{Zel81} Y. B. Zeldovich, Proc. Roy.
Soc. London A, 374, 29 (1981).

\bibitem{Bulle93} B. Dubrulle, Icarus, 106,
59 (1993).

\bibitem{long02} P.-Y. Longaretti, Astrophys. J.,
576, 587 (2002).

\bibitem{Pr25} Z. A. Prandtl, Zs Angew. Math.
Mech., 5, 136 (1925).

\bibitem{BHS96} S. A. Balbus, J. F. Hawley,
and J. M. Stone, Astrophys. J., 467, 76 (1996).

\bibitem{HBW99} J. F. Hawley, S. A. Balbus,
and W.F. Winters, Astrophys. J., 518, 394 (1999).

\bibitem{Spe89} C. G. Speziale, and N. Mac Giolla Mhuiris, Phys. Fluids A, 1,
294(1989).

\end{thebibliography}

\end{document}